\documentclass{article}
\usepackage{cite}
\usepackage{amssymb}
\usepackage{amsmath}
\usepackage{graphicx}
\usepackage{a4}
\usepackage{fullpage}
\usepackage{thm-restate}

\newcommand{\eps}{\varepsilon}

\newcommand{\ceil}[1]{\left\lceil{#1}\right\rceil}
\newcommand{\floor}[1]{\left\lfloor{#1}\right\rfloor}

\newtheorem{lemma}{Lemma}
\newtheorem{theorem}{Theorem}

\newcommand{\qed}{\unskip\nobreak\hfill\ensuremath{\Box}}
\newenvironment{proof}[1][Proof]{\par\noindent\emph{#1. }}{\qed\par\medskip}

\newcommand\drop[1]{}
\newcommand\mtcom[1]{}

\newcommand{\PathPrefix}{\mathsf{PathPrefix}}

\newcommand{\SuffixStateExtend}{\mathsf{SuffixStateExtend}}
\newcommand{\StringStateExtend}{\mathsf{StringStateExtend}}
\newcommand{\PrefixMatch}{\mathsf{PrefixMatch}}
\newcommand{\SubstringMatchExtend}{\mathsf{SubstringMatchExtend}}

\RequirePackage{xcolor}
\newif\ifreviewcomments
\reviewcommentstrue

\title{Faster Regular Expression Matching via Two-Dimensional Tabulation}

\author{Philip Bille \\\texttt{phbi@dtu.dk} \and Inge Li G{\o}rtz \\\texttt{inge@dtu.dk}}

\date{}

\begin{document}
\maketitle
\begin{abstract}
A regular expression specifies a set of strings formed by single characters combined with concatenation, union, and Kleene star operators. Given a regular expression $R$ and a string $Q$, the classic regular expression matching problem is to decide whether $Q$ matches any string specified by $R$. Let $m$ and $n$ be the lengths of $R$ and $Q$, respectively.

 Our main result is a new algorithm that, given a parameter $t \geq m$, solves the problem in 
\[
 O\left(\frac{nm\,\lg\lg(t/m)}{\lg^2(t/m)}+t+n\right)
\]
time and $O(t)$ words of space. In particular, for any fixed $0<\eps<1$, when $m\leq n^{1-\eps}$, we solve the problem in $O(nm \lg \lg n/\lg^2 n + m + n)$ time and $O(n)$ space. Compared to the previous best result [Bille and Thorup, ICALP 2009], this improves the leading term by a factor $\sqrt{\lg n}/\lg\lg n$. 

The key new technique is a two-dimensional tabulation specialized to the
regular expression. 
\end{abstract}


\section{Introduction}
A regular expression $R$ specifies a set of strings formed by characters from an alphabet $\Sigma$ combined with concatenation ($\odot$), union ($\mid$), and Kleene star ($^\ast$) operators. For instance, $(a|(b\odot a))^\ast$ describes the set of strings of $a$s and $b$s such that every $b$ is followed by an $a$. Given a regular expression $R$ and string $Q$, the regular expression matching problem is to decide if $Q$ matches any of the strings specified by $R$. Regular expressions are a fundamental concept in formal language theory, introduced by Kleene in the 1950s~\cite{Kleene1956}, and regular expression matching is a basic tool in computer science for searching and processing text. Standard tools such as \texttt{grep} and \texttt{sed} provide direct support for regular expression matching in files, and the scripting language \texttt{perl}~\cite{Wall1994} is a full programming language that supports regular expression matching easily. Regular expression matching appears in many large-scale data processing applications, such as internet traffic analysis~\cite{JMR2007, YCDLK2006, KDYCT2006}, data mining~\cite{GRS1999}, databases~\cite{LM2001, Murata2001}, computational biology~\cite{NR2003}, and human-computer interaction~\cite{KHDA2012}.

A classic textbook algorithm for regular expression matching, due to Thompson~\cite{Thompson1968} from 1968, constructs and simulates a nondeterministic finite automaton (NFA) $A$ in $O(nm)$ time, where $n$ is the length of $Q$ and $m$ is the total size of $R$, counting each character symbol, occurrence of $\epsilon$, and operator once. The simulation processes $Q$ from left to right and computes a sequence of sets of states $S_0, \ldots, S_n$ such that $S_i$ is the set of states in $A$ to which there is a path from the initial state that matches $Q[1..i]$. In 1985, Galil~\cite{Galil1985} asked whether a faster algorithm existed. Subsequently, a sequence of results~\cite{Myers1992, BFC2008, Bille2006, BT2009, BT2010} improved the $O(nm)$ time bound using either word-level parallelism or tabulation techniques to speed up the automaton simulation. The best known general bounds are either $O(nm\frac{\log w}{w} + n + m\log m)$~\cite{Bille2006, BT2010}, where $w$ is the number of bits in a machine word, or $O(nm\lg\lg t/\lg^{3/2}t+n+m)$ matching time after $O(t)$ preprocessing, where $t$ is the available budget for tables~\cite{BT2009}. 

For a standard logarithmic word size, the latter tabulation algorithm dominates the running time of the word-level parallelism-based algorithm. 

For instance, setting $t=n$ gives $O(nm\lg\lg n/\lg^{3/2}n+n+m)$ time and $O(n+m)$ total space, saving a factor of $\lg^{3/2}n/\lg\lg n$ in the leading $nm$ term of Thompson's algorithm. 

With this choice of $t$, if the alphabet is bounded or $m\leq n^{1-\eps}$ for a fixed $0<\eps<1$, the time bound improves to $O(nm/\lg^{3/2}n+n+m)$~\cite{BT2009}.  

Conversely, several conditional lower bounds suggest that polynomial improvements of the $O(nm)$ bound are not possible. Backurs and Indyk~\cite{BI2016} showed in 2016 that an $O((nm)^{1-\epsilon})$-time algorithm, for any fixed $\epsilon>0$, would refute the orthogonal vectors hypothesis (OVH) (see also Bringmann, Gr{\o}nlund, and
Larsen~\cite{BGL2017}). Abboud and Bringmann~\cite{AB2018} showed that an $O(nm/\lg^{7+\epsilon}n)$-time algorithm, for any fixed $\epsilon>0$, would refute the formula-SAT hypothesis (FSH)\footnote{we note that the recent ETH refutation does not refute OVH or FSH.}. In summary, the best known upper bound saves a factor of $\lg^{3/2}n/\lg\lg n$ over the $O(nm)$ algorithm, whereas the best known conditional lower bound rules out improvements by more than a factor of $\lg^{7+\epsilon}n$, leaving a polylogarithmic gap. 



\subsection{Setup and Results}
Throughout the paper, we assume that $R$ is a regular expression of length $m\geq1$ and $Q$ is a string of length $n\geq1$. We use the convention of $\lg x=\log_2(2+x)$ for $x\geq 0$. As in previous results, we work on a unit-cost word RAM with word length $w \geq \lg (n+m)$, with a standard instruction set including arithmetic, bitwise boolean operations, and shifts. All strings are over an alphabet $\Sigma$ of size $\sigma \geq 2$. We make progress towards closing the polylogarithmic gap with a new improved upper bound achieving the following result.
\begin{restatable}{theorem}{generalalphabettheorem}
\label{thm:general-alphabets}
\label{thm:experimental-general-alphabets}
Let $R$ be a regular expression of length $m$ and let $Q$ be a
string of length $n$ over a word-RAM alphabet. For any integer
$t\geq m$, we can solve regular expression matching in
\[
 O\left(\frac{nm\,\lg\lg(t/m)}{\lg^2(t/m)}+t+n\right)
\]
time and $O(t)$ words of space.
\end{restatable}
In particular, for any fixed $0<\eps<1$, if we set $t=n$ and $m\leq n^{1-\eps}$, we obtain $O(nm\lg\lg n/\lg^2 n+n)$ time and $O(n)$ space. Compared to the previous best bound, we improve the time by a factor of $\sqrt{\lg n}/\lg\lg n$. For bounded alphabets, we avoid the $\lg\lg n$ factor to obtain $O(nm/\lg^2 n+n)$ time, improving the previous best bound by a factor of $\sqrt{\log n}$. 

\subsection{Techniques}\label{subsec:intro-techniques}
We first describe the construction for a constant-size alphabet.
Thompson's simulation performs work for every combination of an
automaton state and a text position. Earlier algorithms partition
the automaton into small micro TNFAs and use table lookups or
word-level operations to process several states at
once~\cite{Myers1992, Bille2006, BFC2008}. These approaches speed up
the automaton dimension but still read one text character at a time.

Bille and Thorup~\cite{BT2009} extend the tabulation to both
dimensions, processing a micro TNFA on a short string at once.
The difficulty is that a matching path can repeatedly traverse
children of a micro TNFA. They summarize each child by the
substrings it accepts and replace complete child traversals by
shortcuts. These summaries contain quadratically many substring
indices, limiting the text dimension to square-root logarithmic
length when the summaries form part of a table index.

Our key idea is to specialize the tables to the given regular
expression. For each short string, we construct the shortcut micro
TNFAs bottom-up during preprocessing, incorporating the behavior
of expanded children into their parents' tables. The child
summaries therefore need not be supplied during matching.
We tabulate four union-preserving functions for extending state
sets and prefix-length sets. Each lookup takes only one such set,
together with a constant number of local states; its table can
be filled from the outputs on singleton sets.

These functions implement a three-phase simulation. A bottom-up
traversal computes the prefix lengths at which an active state
inside each expanded micro TNFA can reach its accepting state.
A top-down traversal computes the prefix lengths at which its
start state is globally reachable. The final phase extends these
prefixes through the remaining suffix and updates the local state
sets. For a table budget $t$, both tabulation dimensions can have
size $\Theta(\lg(t/m))$, giving the squared-logarithmic saving
in the constant-alphabet bound. Unlike the universal tables
of~\cite{BT2009}, our tables depend on the expression.

For general alphabets, we use a second level of decomposition
into larger groups, each containing several micro TNFAs and
having a constant number of interface states. Following the
local alphabet reduction of~\cite[Appendix~A]{BT2009}, a dictionary
encodes the distinct character labels of each group by small
integers and maps all other characters to a common unused code.
We translate each text string once per group and tabulate paths
within that group using its local encoding.

To combine groups, we summarize reachability between interface
states paired with prefix lengths in the current string.
Bottom-up and top-down traversals combine these relations and
propagate the globally reachable interface vertices; local
lookups then recover the final state sets. For micro TNFA size
$x$ and text length $y$, groups of size $X=xy^2$ amortize the
$O(y^2)$ interface computation and the translation cost over their
micro TNFAs. The group alphabets have polynomial size in
$L=\lg(t/m)$, so their codes use $O(\lg L)$ bits. This reduces the
text dimension to $\Theta(L/\lg L)$ while keeping the automaton
dimension at $\Theta(L)$ in the main parameter range, yielding
the additional $\lg\lg(t/m)$ factor for general alphabets.

\subsection{Outline}
Section~\ref{sec:regex} reviews regular expressions and finite
automata. Section~\ref{sec:automata-techniques} presents the
nested decomposition and shortcut micro TNFAs.
Section~\ref{subsec:fast-preprocessing} defines the micro TNFA
functions and shows how to preprocess their tables.
Section~\ref{sec:state-set-transitions} describes the three-phase
state-set simulation and establishes its correctness and
implementation bounds. Section~\ref{subsec:fast-full}
combines these ingredients into the full algorithm for constant alphabets. Finally, 
Section~\ref{sec:general-alphabets} extends the result to general alphabets. 

\subsection{AI Declaration}
The authors used AI (ChatGPT 6) to discuss and verify ideas in an initial phase. We used AI to produce initial drafts, which the authors then heavily edited.

\section{Regular Expressions and Finite Automata}\label{sec:regex}
We briefly review the classical concepts used in the paper. For more
details, see, e.g., Aho et al.~\cite{ASU1986}.

\paragraph{Strings} For a string $q=q[1]\cdots q[\ell]$, we write $q[i..j]$ for
the substring containing positions $i$ through $j$, inclusive,
where $1\leq i\leq j\leq\ell$. We also set
$q[i..i-1]=\epsilon$ for $1\leq i\leq\ell+1$.
In particular, $q[1..0]$ is the empty prefix.

\paragraph{Regular Expressions}
We consider regular expressions over an alphabet $\Sigma$, defined
recursively as follows. Each $\alpha\in\Sigma\cup\{\epsilon\}$ is a
regular expression, where $\epsilon$ denotes the empty string. If $S$
and $T$ are regular expressions, then so are the \emph{concatenation}
$(S)\odot(T)$, the \emph{union} $(S)|(T)$, and the \emph{star} $(S)^*$.
We often omit the concatenation operator $\odot$ when writing regular
expressions. The \emph{language} $L(R)$ generated by a regular expression
$R$ is defined by $L(\alpha)=\{\alpha\}$ for
$\alpha\in\Sigma\cup\{\epsilon\}$,
$L(S\odot T)=L(S)\odot L(T)$,
$L(S|T)=L(S)\cup L(T)$, and
$L(S^*)=\bigcup_{i\geq 0}L(S)^i$.
Here concatenation of languages consists of all concatenations of a
string from the first language and a string from the second,
$L(S)^0=\{\epsilon\}$, and
$L(S)^i=L(S)^{i-1}\odot L(S)$ for $i>0$.

The \emph{parse tree} $T(R)$ is the rooted ordered tree representing
the recursive structure of $R$. Its leaves are labeled by elements
of $\Sigma\cup\{\epsilon\}$, its unary nodes by $*$, and its binary
nodes by $\odot$ or $|$. The length $m$ of $R$ counts each alphabet
symbol, occurrence of $\epsilon$, and operator once; equivalently,
$m$ is the number of nodes of $T(R)$.

\paragraph{Finite Automata}
A \emph{nondeterministic finite automaton} (NFA) is a tuple
$A=(V,E,\Sigma,\Theta,\Phi)$, where $V$ is a finite set of
\emph{states},
$E\subseteq V\times V\times(\Sigma\cup\{\epsilon\})$ is a set of
directed edges called \emph{transitions}, $\Theta\subseteq V$ is a
set of \emph{start states}, and $\Phi\subseteq V$ is a set of
\emph{accepting states}. A transition $(s,s',\alpha)$ goes from $s$
to $s'$ and is labeled by $\alpha$. Thus $A$ is an edge-labeled
directed graph with designated subsets of start and accepting states.
An NFA is a \emph{deterministic finite automaton} (DFA) if it has
exactly one start state, no $\epsilon$-transitions, and at most one
outgoing transition with each character label from each state.

Given a string $q$ and a path $p$ in $A$, we say that $p$ and $q$
\emph{match} if the concatenation of the transition labels along $p$
is $q$. Paths may repeat vertices and transitions, and a path of
length zero matches $\epsilon$. We also say that two paths match
if they match the same string. The set of strings matching some
path from $s$ to $s'$ is denoted by $P_A(s,s')$. For state sets
$S,S'\subseteq V$, define
\[
 P_A(S,S')=\bigcup_{s\in S,\ s'\in S'}P_A(s,s').
\]
We identify a single state with its singleton set when using this
notation. For a state set $S\subseteq V$ and any string $q\in\Sigma^*$,
define the \emph{state-set transition}
\[
 \delta_A(S,q)=\{v\in V\mid\exists s\in S,\ q\in P_A(s,v)\}.
\]
We write $\delta_A(s,q)=\delta_A(\{s\},q)$.
In particular, $\delta_A(S,\epsilon)$ is the epsilon closure of $S$,
including $S$ itself. Paths matching a character may include
$\epsilon$-transitions before and after the character transition.

The automaton $A$ \emph{accepts} a string $Q$ if a path from a state
in $\Theta$ to a state in $\Phi$ matches $Q$; otherwise it
\emph{rejects} $Q$. For $Q$ of length $n$, compute
\[
 S_0=\delta_A(\Theta,\epsilon),\qquad
 S_i=\delta_A(S_{i-1},Q[i]),\quad i=1,\ldots,n.
\]
Then $A$ accepts $Q$ if and only if $\Phi\cap S_n\neq\emptyset$,
including when $Q=\epsilon$.

\begin{figure}[t] 
  \centering \includegraphics[scale=.5]{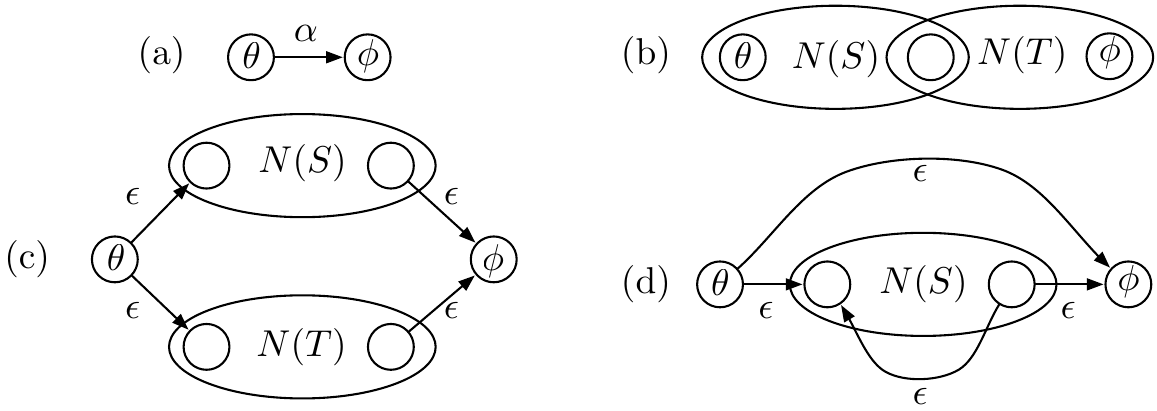}
  \caption{Thompson's recursive NFA construction. The regular
   expression for a character $\alpha \in \Sigma$ corresponds to NFA
   $(a)$. If $S$ and $T$ are regular expressions then $N(ST)$,
   $N(S|T)$, and $N(S^*)$ correspond to NFAs $(b)$, $(c)$, and $(d)$,
   respectively.  In each of these figures, the leftmost state $\theta$
   and rightmost state $\phi$ are the start and accepting states,
   respectively.  For the top recursive calls, these are the start and
   accepting states of the overall automaton. In the recursions indicated,
   e.g., for $N(ST)$ in (b), we take the start state of the
   subautomaton $N(S)$ and identify it with the state immediately to the
  left of $N(S)$ in (b). Similarly the accepting state of $N(S)$ is
   identified with the state immediately to the right of $N(S)$ in
   (b).}

  \label{fig:thompson}
\end{figure}

\paragraph{Thompson's Automata}
Given a regular expression $R$, an NFA accepting precisely $L(R)$
can be obtained by several classic
methods~\cite{MY1960, Glushkov1961, Thompson1968}. In particular,
Thompson~\cite{Thompson1968} gave the construction in
Fig.~\ref{fig:thompson}. For the expression $\epsilon$, use two
distinct states joined by an $\epsilon$-transition. We call an
automaton constructed by these rules a \emph{Thompson NFA} (TNFA)
and denote the TNFA for $R$ by $N(R)$. Each TNFA $A$ has a single
start state $\theta_A$ and a single accepting state $\phi_A$;
thus $\Theta_A=\{\theta_A\}$ and $\Phi_A=\{\phi_A\}$.
We use subscripts to identify the automaton when needed.

A TNFA $N(R)$ for an expression of length $m$ has at most $2m$
states and at most $4m$ transitions, and can be constructed in
$O(m)$ time. Epsilon closure can be computed by a multi-source
breadth-first search restricted to $\epsilon$-transitions. To
compute $\delta_A(S,\alpha)$ for a character $\alpha$, first take
the epsilon closure of $S$, follow all transitions labeled $\alpha$
from that set, and take the epsilon closure of the resulting states.
Each step takes $O(m)$ time for a TNFA. Hence, including construction
and the initial epsilon closure, acceptance of a string of length $n$
can be tested in $O(m(n+1))$ time and $O(m)$ space. For $n\geq1$,
the time bound is $O(nm)$. This is Thompson's
algorithm~\cite{Thompson1968}.

\section{Automata Techniques}\label{sec:automata-techniques}

\subsection{Nested Decomposition}\label{sec:automata-decomposition}
Let $R$ be a regular expression and let $N(R)$ be its TNFA. We define a
\emph{nested decomposition} $\mathcal{AS}$ of $N(R)$ following
Bille~\cite{Bille2006}. A \emph{cluster} is a connected subgraph of the
parse tree $T(R)$, and a \emph{cluster partition} $\mathcal{CS}$ is a
partition of the nodes of $T(R)$ into node-disjoint clusters. Edges
joining different clusters are \emph{external edges}, and all other
edges are \emph{internal edges}. Contracting the internal edges gives
a \emph{macro tree} with one node for each cluster. The parent, child,
root, and leaf relationships between clusters are those of their
corresponding nodes in the macro tree.

Fix a character $\beta \notin \Sigma$. For each cluster $C$ with child
clusters $C_1,\ldots,C_\ell$, insert a \emph{pseudo-node} $p_i$ labeled
$\beta$ on the external edge joining $C$ to $C_i$. Let $T_C$ be the
tree induced by the nodes of $C$ and $p_1,\ldots,p_\ell$. This is a
parse tree for a regular expression $R_C$ over $\Sigma \cup \{\beta\}$.
The TNFA corresponding to $C$ is $A=N(R_C)$, which we call a
\emph{micro TNFA}. The set
\[
    \mathcal{AS}=\{N(R_C)\mid C\in\mathcal{CS}\}
\]
is a \emph{nested decomposition} of $N(R)$. We use the terms parent,
child, root, and leaf for these micro TNFAs as for their clusters.

In a parent micro TNFA $A$, each child micro TNFA $A_i$ is represented by its start
and accepting states $\theta_{A_i}$ and $\phi_{A_i}$ and a
\emph{pseudo-transition} labeled $\beta$ from $\theta_{A_i}$ to
$\phi_{A_i}$. These two states are shared by $A$ and $A_i$.
Recursively replacing the pseudo-transitions by the TNFAs they
represent, identifying the shared states, recovers $N(R)$.

The following lemma is the decomposition lemma of Bille~\cite{Bille2006}
with the binary-tree restriction used by Bille and Thorup~\cite{BT2009}.
The latter follows from Frederickson's tree partition
technique~\cite{frederickson1997}.

\begin{lemma}\label{lem:nested-decomposition}
Given a regular expression $R$ of length $m$ and an integer $x \geq 8$,
we can construct in $O(m)$ time a nested decomposition $\mathcal{AS}$
of $N(R)$ such that
\begin{enumerate}
    \item $|\mathcal{AS}|=O(\ceil{m/x})$,
    \item each micro TNFA in $\mathcal{AS}$ has at most $x$ states, and
    \item each micro TNFA in $\mathcal{AS}$ has at most two children.
\end{enumerate}
In particular, the decomposition tree has maximum degree at most three.
\end{lemma}
\begin{proof}
Set $y=\floor{x/2}-2=\Theta(x)$. Apply Frederickson's tree partition
technique~\cite{frederickson1997} to the binary parse tree $T(R)$,
as in Bille and Thorup~\cite[Section~4]{BT2009}. In $O(m)$ time this
gives a partition into $O(\ceil{m/y})$ node-disjoint clusters, each
with at most $y$ nodes and at most two child clusters. Form the nested
decomposition from this partition as above. Each cluster contributes
one micro TNFA, so $|\mathcal{AS}|=O(\ceil{m/y})=O(\ceil{m/x})$, and the
parent-child relationships are preserved.

For a cluster $C$, the augmented parse tree $T_C$ has at most $y+2$
nodes, including its at most two pseudo-nodes. Thompson's construction
uses at most two states per parse-tree node, so the corresponding micro TNFA
has at most $2(y+2)\leq x$ states. Finally, constructing all augmented
parse trees and their micro TNFAs takes $O(m)$ total time, since the clusters
partition $T(R)$ and there is one pseudo-node per edge of the macro tree.
\end{proof}

For $A\in\mathcal{AS}$ we distinguish the \emph{local} micro TNFA $A$,
in which a child $C$ is represented by its pseudo-transition from
$\theta_C$ to $\phi_C$, from the \emph{expanded} TNFA $\bar A$
obtained by recursively replacing all descendant pseudo-transitions.
The interface states of a child are shared with its parent as
described above. The construction has the following interface
property: transitions from outside $\bar A$ into its interior
pass through its start state $\theta_A$, and paths from its
interior to the outside pass through its accepting state $\phi_A$.
Indeed, the expanded TNFA replaces a single pseudo-transition,
and its start and accepting states are its only interfaces with
the surrounding automaton.
Write $V(A)$ and $V(\bar A)$ for the states of the local and
expanded TNFAs, respectively. For a state set $S$, define its
restrictions by $S|_A=S\cap V(A)$ and
$S|_{\bar A}=S\cap V(\bar A)$. Thus $S|_{\bar A}$ also includes
states in descendants of $A$. We write $S_A=S|_A$ for the local
restriction of the current state set $S$.

\subsection{Shortcut Micro TNFAs}
\label{subsec:fast-automata}
Let $\mathcal{AS}$ be the nested decomposition from
Lemma~\ref{lem:nested-decomposition} with parameter $x$. We use the
local and expanded TNFAs and state-set restrictions defined in
Section~\ref{sec:automata-decomposition}. State sets are stored as
consistent local bitmaps at shared states.

A \emph{string-labeled automaton} is a finite directed multigraph
with designated sets of start and accepting states, whose transitions
are labeled by strings in $\Sigma^*$, including $\epsilon$.
Parallel transitions are allowed. A path matches the concatenation
of its transition labels, and a path of length zero matches $\epsilon$.
The definitions of $P_A$ and $\delta_A$ from Section~\ref{sec:regex}
extend directly to these automata. If a label is stored as a substring
occurrence, its indices specify the stored occurrence; matching uses
the string it spells. We use the path-based definition of $\delta_A$,
since the character-by-character simulation in Section~\ref{sec:regex}
does not directly apply to string-labeled transitions.
We also generalize TNFAs to allow string-labeled transitions in place
of child pseudo-transitions, as in the shortcut micro TNFA defined below.

Fix a nonnegative integer $y$ and a string $q=q[1..\ell]$
of $\ell\leq y$ characters. We represent
prefixes by their lengths $0,\ldots,\ell$, with length $0$ denoting
the empty prefix. A set of prefix lengths is encoded by an
$(\ell+1)$-bit bitmap, and $S_A$ by a bitmap with at most $x$ bits. We use the path-language notation
$P_A$ from Section~\ref{sec:regex}.

For $A\in\mathcal{AS}$ and the string $q$, define the
\emph{shortcut micro TNFA} $A(q)$ following Bille and
Thorup~\cite[Section~4.1]{BT2009}.
It has the local states of $A$, with start state $\theta_A$ and
accepting state $\phi_A$, and retains all local character and
epsilon transitions. For each child $C$, replace the
$\beta$-transition $\theta_C\to\phi_C$ by an edge
\[
 \theta_C\xrightarrow{q[i..j]}\phi_C
 \qquad\text{for each }1\leq i\leq j\leq\ell
 \text{ such that }q[i..j]\in P_{\bar C}(\theta_C,\phi_C).
\]
If $\bar C$ accepts $\epsilon$, also add a single epsilon edge
from $\theta_C$ to $\phi_C$. Thus only child pseudo-transitions
are replaced; no other shortcut edges are added.
A nonempty substring label is stored by its indices $(i,j)$,
without copying the string. Distinct occurrences may give parallel
edges spelling the same string. The empty label is represented once,
using $(1,0)$.

For local states $s,s'$ and $1\leq i\leq\ell+1$,
$i-1\leq j\leq\ell$, we have
\begin{equation}\label{eq:compiled-reachability}
 q[i..j]\in P_{A(q)}(s,s')
 \quad\Longleftrightarrow\quad
 q[i..j]\in P_{\bar A}(s,s').
\end{equation}
Indeed, each traversal of an expanded child along a matching path
can be replaced by its shortcut edge, and each shortcut edge can
be expanded into a matching path through that child. Local
transitions are unchanged. This also handles empty paths and
repeated child traversals.

For every $S\subseteq V(A)$ and every substring $r$
of $q$, including $\epsilon$, the shortcut micro TNFA satisfies
\[
 \delta_{A(q)}(S,r)=V(A)\cap\delta_{\bar A}(S,r).
\]
Indeed, every edge of $A(q)$ represents a path in $\bar A$, so
expanding the edges of a matching path proves one inclusion.
Conversely, replacing each complete child traversal in a path
through $\bar A$ by a shortcut gives a matching path in $A(q)$,
proving the other inclusion.

\begin{lemma}\label{lem:shortcut-graph}
Let $A$ belong to a nested decomposition satisfying
Lemma~\ref{lem:nested-decomposition} with parameter $x$,
and let $q$ have length $\ell$. The shortcut micro TNFA $A(q)$ has
at most $x$ states and $O(x+(\ell+1)^2)$ transitions.
Given the shortcut micro TNFAs $C(q)$ for the children of $A$, it can
be constructed in $O((\ell+1)^2(x+\ell+1))$ time and
$O(x(\ell+1)+(\ell+1)^2)$ words of workspace, including the output.
If the accepted substring indices for each child have already
been computed, construction takes $O(x+(\ell+1)^2)$ time and space.
\end{lemma}
\begin{proof}
The local micro TNFA has $O(x)$ transitions. Each of its at most two
children contributes at most $\ell(\ell+1)/2$ nonempty substring
edges and one epsilon edge. This gives the size bound and the
construction bound when the accepted substring indices are known.

To obtain these indices from a child shortcut micro TNFA $C(q)$,
temporarily make a copy $(s,p)$ of each state $s$ of $C$ for
each prefix length $0\leq p\leq\ell$. For each local epsilon
transition or empty shortcut $s\to s'$, add
$(s,p)\to(s',p)$ for every $0\leq p\leq\ell$. For each
$0\leq p<\ell$ and each local character transition $s\to s'$
matching $q[p+1]$, add $(s,p)\to(s',p+1)$.

For each nonempty shortcut labeled by the occurrence $q[i..j]$,
add an edge from $(s,i-1)$ to $(s',j)$.
There are $O(x(\ell+1))$ vertices and
$O(x(\ell+1)+(\ell+1)^2)$ edges.
A shortcut for every accepted nonempty occurrence is present.
If a shortcut label matches another occurrence of the same string
in $q$, the construction also contains the shortcut stored for
that occurrence. We can therefore use the shortcut corresponding
to the positions currently being read. Replacing child traversals
then shows that $(s,i-1)$ reaches
$(s',j)$ exactly when $q[i..j]\in P_{C(q)}(s,s')$.

For every $0\leq p\leq\ell$, perform a graph search from
$(\theta_C,p)$, marking its source as reachable. Each reachable
$(\phi_C,j)$ with $j>p$ gives the accepted substring $q[p+1..j]$.
Reachability of $(\phi_C,p)$ detects acceptance of $\epsilon$,
which is recorded once. These $\ell+1$ searches take
$O((\ell+1)^2(x+\ell+1))$ time and use the stated workspace.
Repeat for each child and insert the resulting shortcuts in $A$.
\end{proof}

\section{Preprocessing for State-Set Transitions}\label{subsec:fast-preprocessing}

We now show how to efficiently preprocess micro TNFAs in the nested decomposition and small substrings together. We construct the shortcut micro TNFAs during preprocessing and tabulate
four functions on them, called \emph{micro TNFA functions}, allowing
the simulation to evaluate state-set transitions by table lookup. We show how to tabulate these efficiently and that they suffice to implement state-set transitions in Section~\ref{sec:state-set-transitions}. 

Let $\mathcal{AS}$ be a nested decomposition with parameter $x$. Let $A \in \mathcal{AS}$ and let $q$ be a string of length $\ell \leq y$. For $I\subseteq\{0,\ldots,\ell\}$, $S\subseteq V(A)$, and
$s,s'\in V(A)$, define
\[
\begin{aligned}
 \SuffixStateExtend_{A,q}(I,s)
   &=\bigcup_{i\in I}\delta_{A(q)}(s,q[i+1..\ell]),\\
 \StringStateExtend_{A,q}(S)
   &=\delta_{A(q)}(S,q[1..\ell]),\\
 \PrefixMatch_{A,q}(S,s')
   &=\{j \mid s'\in\delta_{A(q)}(S,q[1..j])\},\\
 \SubstringMatchExtend_{A,q}(I,s,s')
   &=\{j \mid \exists i\in I,\ i\leq j,\ 
          s'\in\delta_{A(q)}(s,q[i+1..j])\}.
\end{aligned}
\]
In the last two definitions, $j$ ranges over $0,\ldots,\ell$.
The set $I$ contains lengths of prefixes already consumed, so an
extension from length $i$ to length $j$ matches $q[i+1..j]$.
In particular, $i=j$ allows an empty extension.
The first two functions return local state sets, and the last two return
sets of prefix lengths. For fixed state arguments,
$\SuffixStateExtend$ and $\SubstringMatchExtend$ preserve unions
in $I$, and $\StringStateExtend$ and $\PrefixMatch$ preserve unions
in $S$. All four functions map the empty set to the empty set.

Intuitively, $\SuffixStateExtend$ starts at $s$ after any prefix
length in $I$ has been consumed and returns all states reachable
after reading the remaining suffix of $q$.
$\StringStateExtend$ returns all states reachable from $S$ after
reading the entire string $q$.
$\PrefixMatch$ returns the lengths of prefixes of $q$ that take
some state in $S$ to $s'$.
Finally, $\SubstringMatchExtend$ starts at $s$ after any prefix
length in $I$ and returns all prefix lengths at which $s'$ can be
reached by reading the intervening substring.
All paths are taken in the shortcut micro TNFA $A(q)$.

\subsection{Encoding and Tabulation}
We now show how to encode and tabulate the micro TNFA functions efficiently. More precisely, we will represent each function as a lookup table indexed by its argument and store the output in the corresponding entry. We encode a local state set (parameter $S$) as a bitstring of at most $x$ bits, and a set of indexes (parameter $I$) as a bitstring of length $\ell+1$. We encode each local state (parameters $s$ and $s'$) as $\ceil{\lg x}$-bit integers. Similarly, we encode state set output for $\SuffixStateExtend$ and $\StringStateExtend$ as a bitstring of at most $x$ bits, and the set of indexes output for $\PrefixMatch$ and $\SubstringMatchExtend$ as bitstrings of length $\ell +1$. We obtain the following tabulation. 

\begin{lemma}\label{lem:micro-tabulation}

Let $\mathcal{AS}$ be the nested decomposition from
Lemma~\ref{lem:nested-decomposition} for a regular expression of
length $m$, with parameter $x$, and let $\Sigma$ be a constant alphabet of size
$\sigma\geq 2$. Assume that the encoded input and output masks fit
in $O(1)$ machine words and that the tables are addressable.
Using the above encoding, we can tabulate the four
micro TNFA functions for every $A\in\mathcal{AS}$, every
$q\in\Sigma^{\leq y}$, and all choices of local state arguments in
\[
 O\left(m+\ceil{m/x}\sigma^y(x2^x+x^2 2^y)\right)
\]
time and words of space. Each evaluation takes $O(1)$ time.
\end{lemma}
\begin{proof}
Fix $(A,q)$ and let $a=|V(A)|\leq x$ and $\ell=|q|$.
For each choice of state arguments, store a separate table indexed
by the set argument. The numbers of entries for
$\SuffixStateExtend$, $\StringStateExtend$, $\PrefixMatch$, and
$\SubstringMatchExtend$ are, respectively,
\[
 a2^{\ell+1},\qquad 2^a,\qquad a2^a,\qquad a^2 2^{\ell+1}.
\]
Their sum is $O(x2^x+x^2 2^y)$, and each entry occupies $O(1)$
words. The encoded state arguments select the appropriate table,
and the encoded set argument selects its entry.

To construct the tables, build the layered graph from
Lemma~\ref{lem:shortcut-graph} for $A(q)$ and perform a graph
search from every $(s,i)$, where $s\in V(A)$ and
$0\leq i\leq\ell$. These searches provide all singleton images:
from $(s,i)$ record reachable states at length $\ell$ and,
for each $s'$, the lengths at which $s'$ is reachable.
The searches and recording take
$O(x(\ell+1)^2(x+\ell+1))$ time.
For each fixed choice of state arguments, fill the table by
preservation of unions: set $f(\emptyset)=\emptyset$ and use
\[
 f(X)=f(X\setminus\{b\})\cup f(\{b\})
\]
for any set bit $b$ of a nonzero input bitmap $X$.
Enumerating bitmaps by their highest set bit gives constant work
per entry without requiring a bit-search instruction.

For each $q$, construct the shortcut micro TNFAs bottom-up using
Lemma~\ref{lem:shortcut-graph}. The construction, searches, and
temporary storage per micro TNFA are polynomial in $x+y+1$ and are
absorbed by $O(x2^x+x^2 2^y)$: for $x\geq 1$ and $y\geq 0$,
every fixed-degree polynomial in $x+y+1$ is
$O(2^x+2^y)$. By Lemma~\ref{lem:nested-decomposition}, there
are $O(\ceil{m/x})$ micro TNFAs. Summing over these micro TNFAs and
$O(\sigma^y)$ strings, and allowing $O(m)$ time and space for
the decomposition, proves the bounds.
\end{proof}

\section{Computing State-Set Transitions}\label{sec:state-set-transitions}
Let $N=N(R)$. We now show how to efficiently implement state-set transitions using the preprocessing from the previous section. The algorithm proceeds in the following three phases. The first two phases traverse the nested decomposition to compute auxiliary information, which we then use in the final phase to perform the state-set transition. 

\subsection{Phase 1: Computing Path Prefixes in $\overline{A}$ to $\phi_A$}
\label{subsec:fast-phase1}
Let $S$ be the global state set immediately before reading $q$.
For every $A\in\mathcal{AS}$ define
\[
 \PathPrefix_{A,q}^\phi=
 \{j \mid 0\leq j\leq\ell,\;
       q[1..j]\in P_{\bar A}(S|_{\bar A},\phi_A)\}.
\]
Intuitively, $\PathPrefix_{A,q}^\phi$ records how many characters of
$q$ can be read before reaching the accepting state of $A$, starting from
an active state inside $\bar A$ and staying within $\bar A$.

We compute $\PathPrefix_{A,q}^\phi$ in a bottom-up traversal of $\mathcal{AS}$ using the micro TNFA functions as follows. 
\begin{equation}\label{eq:fast-phase1}
\begin{split}
 \PathPrefix_{A,q}^\phi
 ={}&\PrefixMatch_{A,q}(S_A,\phi_A)\\
 &{}\cup\bigcup_{C\text{ child of }A}
      \SubstringMatchExtend_{A,q}(\PathPrefix_{C,q}^\phi,\phi_C,\phi_A).
\end{split}
\end{equation}

Intuitively, the first term accounts for paths starting at an
active local state of $A$ and reaching $\phi_A$.
Each term in the union accounts for paths starting inside a child
$C$: $\PathPrefix_{C,q}^\phi$ gives the prefix lengths at which
they can reach $\phi_C$, and $\SubstringMatchExtend$ extends
these paths from $\phi_C$ to $\phi_A$ inside $\bar A$.
We formally prove~\eqref{eq:fast-phase1} in
Lemma~\ref{lem:phase-correctness}.

\subsection{Phase 2: Computing Path Prefixes in $N(R)$ to $\theta_A$}
\label{subsec:fast-phase2}
We next compute the prefix lengths at which the start state of $A$
is reachable from $S$ in the full TNFA $N$.
\[
 \PathPrefix_{A,q}^\theta=
 \{j \mid 0\leq j\leq\ell,\;
           q[1..j]\in P_N(S,\theta_A)\}.
\]
Intuitively, $\PathPrefix_{A,q}^\theta$ records how many characters of
$q$ can be read before reaching the start state of $A$, starting
from any active state in the full TNFA $N$.
For the root $A_0$ of $\mathcal{AS}$, its start state $\theta_{A_0}$
is the start state of $N$; under the usual Thompson construction
this state has no incoming transitions. Thus
$\PathPrefix_{A_0,q}^\theta=\{0\}$ if $\theta_{A_0}\in S$, and
$\PathPrefix_{A_0,q}^\theta=\emptyset$ otherwise.

Let $A$ be a child of $B$. A path from $S$ to $\theta_A$ can
originate in a state local to $B$, enter $\bar B$ via
$\theta_B$, or originate in a child $C$ of $B$ and first leave
that child at $\phi_C$. It follows that
\begin{equation}\label{eq:fast-phase2}
\begin{split}
 \PathPrefix_{A,q}^\theta
 ={}&\PrefixMatch_{B,q}(S_B,\theta_A)\\
    &{}\cup \SubstringMatchExtend_{B,q}(\PathPrefix_{B,q}^\theta,\theta_B,\theta_A)\\
    &{}\cup\bigcup_{C\text{ child of }B}
          \SubstringMatchExtend_{B,q}(\PathPrefix_{C,q}^\phi,\phi_C,\theta_A).
\end{split}
\end{equation}
Intuitively, the first term accounts for paths starting at an
active local state of $B$ and reaching $\theta_A$.
The second term extends paths that have reached $\theta_B$,
using $\PathPrefix_{B,q}^\theta$ to specify the consumed prefix
lengths. Each term in the final union extends paths that start
inside a child $C$ and reach its accepting state $\phi_C$, using
$\PathPrefix_{C,q}^\phi$ for these prefix lengths.
In each case, the extension to $\theta_A$ stays inside $\bar B$.
We formally prove~\eqref{eq:fast-phase2} in
Lemma~\ref{lem:phase-correctness}.

The accepting-state prefix sets of \emph{both} children of $B$ may occur,
including the destination child $A$. After completing Phase~1,
we compute $\PathPrefix_{A,q}^\theta$ using~\eqref{eq:fast-phase2}
in a top-down traversal of the nested decomposition, visiting
parents before their children.

\subsection{Phase 3: Updating State Sets}
\label{subsec:fast-phase3}
Finally, let $S'=\delta_N(S,q)$. The states of $A$ reachable
after all $\ell$ characters are
\begin{equation}\label{eq:fast-phase3}
\begin{split}
 S'_A
 ={}& \StringStateExtend_{A,q}(S_A)\\
    &{}\cup \SuffixStateExtend_{A,q}(\PathPrefix_{A,q}^\theta,\theta_A)\\
    &{}\cup\bigcup_{C\text{ child of }A}
               \SuffixStateExtend_{A,q}(\PathPrefix_{C,q}^\phi,\phi_C).
\end{split}
\end{equation}
Intuitively, the first term accounts for paths starting at an
active local state of $A$ and reading all of $q$.
The second term uses $\PathPrefix_{A,q}^\theta$ to extend paths
from the start state $\theta_A$ through the remaining suffix.
Each term in the final union similarly uses
$\PathPrefix_{C,q}^\phi$ to extend paths from a child's accepting state
$\phi_C$. In each case, the extension stays inside $\bar A$
and returns the local states reachable after all of $q$.
We formally prove~\eqref{eq:fast-phase3} in
Lemma~\ref{lem:phase-correctness}.
Each computed $S'_A$ is the restriction of $\delta_N(S,q)$ to
$V(A)$, so the copies of shared states agree. These local state
sets are used for the next substring.

\subsection{Correctness}\label{subsec:phase-correctness}
\begin{lemma}\label{lem:phase-correctness}
Let $S$ be a state set of $N$ and let $q=q[1..\ell]$.
Phase~1 computes $\PathPrefix_{A,q}^\phi$ as defined in
Section~\ref{subsec:fast-phase1} for every $A\in\mathcal{AS}$.
Phase~2 computes $\PathPrefix_{A,q}^\theta$ as defined in
Section~\ref{subsec:fast-phase2}. Phase~3 computes
\[
 S'_A=V(A)\cap\delta_N(S,q)
 \qquad\text{for every }A\in\mathcal{AS}.
\]
Consequently, the three phases implement the state-set transition
$\delta_N(S,q)$.
\end{lemma}
\begin{proof}
By the shortcut property~\eqref{eq:compiled-reachability},
a path inside $\bar A$ between local states, matching a substring
of $q$, can be replaced by a matching path in $A(q)$, and conversely.
We use this property to justify the three recurrences.

For Phase~1, proceed by induction from leaves to the root.
Consider a path inside $\bar A$ from $S|_{\bar A}$ to $\phi_A$
matching $q[1..j]$. If its initial state is local to $A$, the
shortcut property places $j$ in the first term of
\eqref{eq:fast-phase1}. Otherwise it starts inside a child $C$.
Take its first visit to that child's accepting state $\phi_C$, after consuming
$i$ characters. The initial segment stays inside $\bar C$, so
$i\in\PathPrefix_{C,q}^\phi$ by induction. The remaining segment
stays inside $\bar A$ and matches $q[i+1..j]$, giving the
corresponding child term. Conversely, the first term expands
to a path of the required kind, and each child term concatenates
such an initial path in $\bar C$ with a path in $\bar A$.
This proves the recurrence, including the leaf case.

For Phase~2, proceed by induction from the root to the leaves.
The root initialization is correct because the start state of $N$
has no incoming transitions. Let $A$ be a child of $B$ and
consider a path from $S$ to $\theta_A$ matching $q[1..j]$.
If the path is not entirely inside $\bar B$, split it at its
last entry into $\bar B$, through its start state $\theta_B$
by the interface property from Section~\ref{sec:automata-decomposition}.
By induction, the consumed prefix length belongs to
$\PathPrefix_{B,q}^\theta$, and the remaining path gives the second
term of~\eqref{eq:fast-phase2}. If the path stays inside $\bar B$,
it either starts locally in $B$, giving the first term, or starts
inside a child $C$ of $B$. In the latter case split at its first
visit to the accepting state $\phi_C$ of $C$; Phase~1 supplies the consumed prefix length in
$\PathPrefix_{C,q}^\phi$, giving a term in the final union.
Conversely, each term concatenates a valid prefix path with a
path inside $\bar B$ ending at $\theta_A$. Thus the recurrence
computes exactly the required prefix lengths for reaching the start state.

Finally, consider a path from $S$ to a local state $s'$ of $A$
matching all of $q$. If it leaves or starts outside $\bar A$,
split at its last entry through the start state $\theta_A$,
using the same interface property.
Phase~2 supplies the prefix length, and the remaining suffix
gives the second term of~\eqref{eq:fast-phase3}.
Otherwise the path stays inside $\bar A$. It either starts
locally, giving the first term, or starts inside a child $C$;
splitting at its first visit to the child's accepting state gives a term in the final union
by Phase~1. Conversely, each term expands to a path from $S$
to its output state matching $q$. Hence the output is exactly
$V(A)\cap\delta_N(S,q)$. These local sets cover all states and
agree at shared interfaces, so together they represent
$\delta_N(S,q)$.

All splits are at states, and may consume zero characters.
The argument therefore also includes epsilon paths, empty
substrings, and repeated child traversals.
\end{proof}

\subsection{Encoding and Analysis}\label{subsec:phase-encoding}
We use the encoding of the micro TNFA functions from
Section~\ref{subsec:fast-preprocessing}. Number the local states of
each $A$ and encode $S_A$ and $S'_A$ as bitstrings of at most $x$
bits. Encode $\PathPrefix_{A,q}^\phi$ and
$\PathPrefix_{A,q}^\theta$ as bitstrings of length $\ell+1$,
where bit $j$ indicates membership of prefix length $j$.
The shared states retain their local numbers in each micro TNFA;
parent-child pointers store these numbers for the interface states.
Each state argument is therefore available as a
$\ceil{\lg x}$-bit integer.

For each fixed $(A,q)$, the tables from
Lemma~\ref{lem:micro-tabulation} take the encoded state arguments
and set argument and return the encoded output. We evaluate each
term of~\eqref{eq:fast-phase1}--\eqref{eq:fast-phase3} by one
lookup and implement each union by bitwise OR.
The prefix-length encodings are the same for all micro TNFAs processing
$q$, so a child's prefix mask can be used directly in its parent's
tables. The output of Phase~3 uses the same local-state encoding
as the input of Phase~1 for the next string.
We obtain the following implementation bound.

\begin{lemma}\label{lem:phase-implementation}
Given the tables of Lemma~\ref{lem:micro-tabulation}, a string
$q$ of length $\ell\leq y$ together with its table index, and the
encoded local restrictions of $S$, the three phases compute the
encoded local restrictions of $\delta_N(S,q)$ in
$O(\ceil{m/x})$ time using $O(m)$ space in addition to the tables, provided the encoded inputs and output fit in $O(1)$ words and the tables are addressable. 
\end{lemma}
\begin{proof}
Each micro TNFA has at most two children. Phase~1 uses at most three
lookups per micro TNFA, Phase~2 at most four per parent-child pair, and
Phase~3 at most four per micro TNFA. Each lookup and each bitwise union
takes $O(1)$ time. A bottom-up traversal evaluates Phase~1,
a top-down traversal evaluates Phase~2, and one further traversal
evaluates Phase~3. By Lemma~\ref{lem:nested-decomposition}, the
decomposition contains $O(\ceil{m/x})$ micro TNFAs, proving the time bound.
For each micro TNFA we store the current and updated local state masks
and the two prefix masks, using $O(1)$ words. We retain the current
state masks until Phase~3 finishes. Correctness follows from
Lemma~\ref{lem:phase-correctness}, which also ensures that the
updated masks agree at shared states.
\end{proof}

\section{The Full Algorithm}\label{subsec:fast-full}
We now combine the tabulation of the micro TNFA functions with the
three-phase state-set simulation to obtain the main result for
constant-size alphabets. We describe the full matching algorithm
and analyze its time and space bounds, proving
Theorem~\ref{thm:pattern-specific-tabulation}.

\begin{theorem}
\label{thm:pattern-specific-tabulation}
Let $R$ be a regular expression of length $m$ and let $Q$ be a
string of length $n$ over a constant-size alphabet. For any integer
$t\geq m$, we can solve regular expression matching in
\[
 O\left(\frac{nm}{\lg^2(t/m)}+t+n\right)
\]
time and $O(t)$ words of space.
\end{theorem}

\begin{proof}
Fix an integer $t\geq m$. Put $L=\lg(t/m)$ and define
\[
 c=\frac{1}{4(3+\lg\sigma)},\qquad
 L_0=32(3+\lg\sigma)=\frac{8}{c},\qquad b=\floor{cL}.
\]
Then $L>L_0$ implies $b\geq8$. We distinguish the following
three cases.

\paragraph{Case 1: Constant $L$.}
If $L\leq L_0$, we simply apply Thompson's algorithm using $O(nm)$ time and $O(m)$ space. Since $L$ is constant, this matches the time and space bounds.

\paragraph{Case 2: A single micro TNFA.}
Suppose $L>L_0$ and $m<b$. The expression is small enough that
we can tabulate transitions for the entire TNFA, without a
nested decomposition into smaller pieces.
Take $N=N(R)$ as a single micro TNFA with 
$x=2m$ states, since Thompson's construction has at most $2m$ states,
and set $y=b$. Using the table construction in the proof of
Lemma~\ref{lem:micro-tabulation}, we tabulate only
$\StringStateExtend_{N,q}$ for every $q\in\Sigma^{\leq y}$.
Since $N$ has no children, this table maps $S$ directly to
$\delta_N(S,q)$. We initialize $S=\delta_N(\theta_N,\epsilon)$ and partition $Q$ into
consecutive strings of length $y$, except possibly the shorter
final string. For each string $q$, we update
$S=\StringStateExtend_{N,q}(S)$ by one lookup. Finally, we accept if and only if $\phi_N\in S$ after processing all strings.

The table construction takes
\[
 O\left(m+\sigma^b(x2^x+x^2 2^b)\right)
   =O\left(m+b^2 2^{(2+\lg\sigma)b}\right)
\]
time and space, since $x=2m<2b$.
For $b\geq8$, we have $b^2\leq2^b$, so the choice of $c$ gives
\begin{equation}\label{eq:fast-table-factor}
 b^2 2^{(2+\lg\sigma)b}
 \leq 2^{(3+\lg\sigma)b}
 \leq 2^{L/4}.
\end{equation}
Thus preprocessing takes $O(m+2^{L/4})=O(t)$ time and space.
Each string requires one lookup, so matching takes $O(m+n)$
time and $O(m)$ space beyond the tables. Hence this case
satisfies the claimed time and space bounds.

\paragraph{Case 3: The general case.}
Suppose $L>L_0$ and $m\geq b$. We use the nested decomposition
with $x=y=b$, giving $O(m/b)$ micro TNFAs.
Construct $N=N(R)$ and tabulate the four micro TNFA functions
for every micro TNFA $A$ and every $q\in\Sigma^{\leq y}$ using
Lemma~\ref{lem:micro-tabulation}.

In this case, the micro TNFAs have at most $x=b$ states,
and their number is $O(m/b)$. Number the alphabet symbols by
$0,\ldots,\sigma-1$ and index each string's tables by its length
and base-$\sigma$ value. For each micro TNFA, a table indexed
by these two values selects the corresponding tables.

Initialize $S=\delta_N(\theta_N,\epsilon)$ and its local
restrictions. Partition $Q$ into consecutive strings of length
$y$, except possibly the shorter final string.
For each string $q$, compute its table index and update $S$
using the three phases of
Section~\ref{subsec:fast-phase1}--\ref{subsec:fast-phase3}.
Accept if and only if $\phi_N\in S$ after all strings have been
processed. By Lemma~\ref{lem:phase-correctness}, each update
computes $\delta_N(S,q)$.
Thus the final test decides whether $Q\in L(R)$.

By Lemma~\ref{lem:micro-tabulation}, the preprocessing time and
space, including the tables, are bounded by
\[
 O\left(m+(1+m/b)b^2 2^{(2+\lg\sigma)b}\right).
\]
By~\eqref{eq:fast-table-factor}, the factor
$b^2 2^{(2+\lg\sigma)b}$ is at most $2^{L/4}$.
Since $1+m/b=O(m)$ and $2^L=2+t/m$, preprocessing therefore takes
\[
 O(m2^{L/4})=O(t).
\]
Initialization takes $O(m)$ time, and encoding the input takes
$O(n)$ time. Lemma~\ref{lem:phase-implementation} gives
$O(m/b)=O(1+m/b)$ time per string, including the final short string.
The matching time is therefore
\[
 O\left(\ceil{n/b}(1+m/b)+m+n\right)
   = O\left(\frac{nm}{L^2}+m+n\right).
\]
The ceiling contributes only $O(1+m/b)$ extra work, so this bound
also applies when $n<b$. The simulation uses $O(m)$ words
in addition to the tables. Adding preprocessing and using $t\geq m$
completes the proof for Case~3 and hence for all three cases.
\end{proof}

For fixed $\eps>0$ and $t\geq m^{1+\eps}$, we have
$L=\Theta(\lg t)$, so the time bound becomes
$O(nm/\lg^2 t+t+n)$. At $t=m$, $L$ is constant and ordinary
Thompson simulation gives $O(nm+m+n)$ time and $O(m)$ space.

\section{General Alphabets}\label{sec:general-alphabets}
We extend Theorem~\ref{thm:pattern-specific-tabulation} to general alphabets,
proving Theorem~\ref{thm:general-alphabets}, which we restate below.

\generalalphabettheorem*

\subsection{Grouping and alphabet reduction}

Use Lemma~\ref{lem:nested-decomposition} twice. First apply it to
$R$ with a parameter $X\geq x\geq8$ to form groups. Each group
$G$ is the local TNFA of a cluster in this first decomposition.
It has at most $X$ states and at most two children in the group tree.
Call the states shared with its parent or children its
\emph{interface states}. Each group has at most six such states.
We retain the states of the pseudo-transitions to child groups and
exclude those transitions from matching inside $G$. Next decompose
the expression for each group with parameter $x$. We call the pieces
in this second decomposition micro TNFAs. The first decomposition's
pseudo-label is an ordinary symbol in the group expression and
never occurs in an encoded text block. The second decomposition
uses its own pseudo-label. If $m<X$, we may instead use $N(R)$
as a single group with no interfaces.

As in Bille and Thorup~\cite[Appendix~A]{BT2009}, assign the distinct
character labels in each group distinct positive integer codes.
The dictionary for $G$ maps every other text character to $0$.
Reserve a separate code for the excluded pseudo-transitions.
The resulting alphabet has size $O(X+1)$. Let $q_G$ be the
encoding of a text block $q$ in $G$. This encoding preserves every
character test in the group. The groups use the same prefix lengths
$0,\ldots,|q|$ in their encodings, so a reachability relation on
interface states and prefix lengths can be passed between groups.

\subsection{Tables within a group}

For each group $G$ and each encoded block $q_G$ of length $\ell\leq y$,
tabulate the micro TNFA functions for the group expression, as in
Lemma~\ref{lem:micro-tabulation}. The alphabet size enters that
construction through the enumeration of encoded blocks. For a
fixed $q_G$, the construction works on the same layered graphs and
can use all the descendants in the group's second decomposition.

Also consider the layered graph of the entire local group on
$q_G$. Its vertices are pairs $(v,i)$ for a state $v$ of $G$ and a
prefix length $0\leq i\leq\ell$. Epsilon transitions preserve the
prefix length and matching character transitions advance it by one.
For every interface state $v$ of $G$, call all $(v,i)$
\emph{interface vertices}. Store the reachability relation between
all interface vertices within the local group, including paths
of length zero. For each micro TNFA in $G$, store two further kinds
of tables:
\begin{itemize}
    \item Given a local state mask at prefix length $0$, return
    the interface vertices reachable from that mask within $G$.
    \item For each interface state $v$, given a mask of prefix
    lengths $i$, return the local state mask at prefix length
    $\ell$ reachable from the selected vertices $(v,i)$ within $G$.
\end{itemize}
For each table, graph searches give the outputs for singleton
inputs. All other outputs are unions of the corresponding
singleton outputs. Thus the time and space to construct these
tables, the interface relation, and the micro TNFA functions
for one group and one encoded block are bounded by
$(2^x+2^y)\operatorname{poly}(X+y)$.
In particular, the tables for a group depend only on its
own character encoding and local transitions.

\subsection{State-set transitions across groups}

Let $S$ be the state set before reading $q$. We compute the
state set after reading $q$ in three steps.

\paragraph{Bottom-up traversal.}
For each group $G$, construct a graph on its interface vertices.
Start with the stored local reachability relation. For each child
$C$, insert the relation between the vertices of its interface
with $G$ that is realized by paths inside the subtree of $C$.
These vertices already occur in the graph of $G$, since the
groups share the corresponding states. Compute the transitive
closure of the resulting graph. By induction, this closure gives
exactly the reachability between interface vertices of $G$ by paths
in the subtree of $G$. A path in that subtree can be split at
group interfaces into local paths and excursions into child
subtrees. Conversely, each inserted edge represents a path in
the subtree. Restrict the closure to the interface with the parent
to obtain the relation passed up to that parent.

In the same traversal, compute the interface vertices reachable
from $S$ at prefix length $0$ using only the subtree of $G$.
Take the union of the masks returned by the first kind of table
for the micro TNFAs in $G$. Add the reachable vertices reported
by each child at its interface with $G$. Close the resulting
set under the computed relation. This gives exactly the desired
set by splitting a path at the first interface vertex it visits.
Pass its restriction to the parent interface up to the parent.

\paragraph{Top-down traversal.}
At the root, the set computed bottom-up already gives global
reachability. At a nonroot group $G$, take the union of its
bottom-up set and the globally reachable vertices on its
interface with its parent. Close this union under the relation
computed for $G$. This gives the globally reachable interface
vertices of $G$. A path from $S$ either stays within the subtree
of $G$ or has a last entry into that subtree through the interface
with the parent. Each vertex obtained by the closure is reachable
by concatenating valid paths.

\paragraph{Local state sets.}
For each group, run the three phases of
Section~\ref{sec:state-set-transitions} on the group expression
with $q_G$ and the initial state set restricted to that group.
This finds the final states reachable by paths entirely within
the local group. For each micro TNFA, take the union of its
resulting local mask and the masks returned by the second kind of
table for the globally reachable interface vertices.
Any path that reaches a final state in the group either stays
within that group or has a last entry through an interface vertex.
Conversely, each local mask represents valid paths from $S$.
The resulting state sets agree at shared states and represent
$\delta_N(S,q)$.

Each interface graph has $O(y)$ vertices. Its adjacency and
reachability matrices can be stored as $O(y)$ rows of $O(y)$ bits
each. To compute the closure, process the vertices as intermediate
vertices in turn. For a vertex $v$, OR its row into each row whose
entry for $v$ is set. With each row in $O(1)$ words, this takes
$O(y^2)$ time. The restrictions and set closures in both traversals
fit within the same bound. Apart from this work, the time per
group per block is $O(y)$ for encoding and $O(1)$ per micro TNFA
for table lookups and the local state-set computation. A single
group with no interfaces needs only the local computation.

\subsection{Analysis}

\begin{proof}[Proof of Theorem~\ref{thm:experimental-general-alphabets}]

Let $L=\lg(t/m)$ and $M=\min\{m,L\}$. If $M$ is bounded, use
ordinary Thompson simulation in $O(nm+m+n)$ time and $O(m)$
space. If $L$ is bounded, $\lg L/L^2=\Theta(1)$. If $m$ is
bounded, the simulation takes $O(m+n)$ time.

Otherwise set
\[
 x=\floor{aM},\qquad y=\floor{bM/\lg M},\qquad X=xy^2
\]
for sufficiently small positive constants $a,b$. Take the
threshold for $M$ large enough that $8\leq x\leq m$ and
$1\leq y\leq x$. The group expressions have size $O(X)$,
including any pseudo-labels. If $X>m$, use a single group.
If $X\leq m$, there are $O(m/X)$ groups, and the sum of their
expression sizes is $O(m)$. In both cases the total number of
micro TNFAs is $O(m/x)$.

The number of encodings per group is bounded by
$(y+1)(C(X+1))^y$ for a constant $C$. The cost of all tables
for a group is therefore bounded by
\[
 (y+1)(C(X+1))^y(2^x+2^y)\operatorname{poly}(X+y).
\]
Its base-two logarithm is at most
$x+O(y\lg(X+1)+\lg(X+y))$. Since $X=O(M^3)$, the constants
$a,b$ can be chosen so that this is at most $M/2$ for sufficiently
large $M$. With at most $O(m)$ groups and $2^M\leq 2^L=2+t/m$,
preprocessing the tables costs $O(t)$ time and space.
The same slack as in Section~\ref{subsec:fast-full}
ensures the tables are addressable, and the masks, rows, and block
encodings fit in $O(1)$ words. Constructing the group dictionaries
takes $O(m\lg X)$ time and $O(m)$ space by the static dictionary
construction used in Bille and Thorup~\cite[Appendix~A]{BT2009}.
This includes gathering the distinct labels by sorting within
each group. Since $\lg X=O(\lg M)$, the time is absorbed by $O(t)$.

In the case of multiple groups, the total time for encoding and
interface work per block is $O((m/X)y^2)=O(m/x)$. In the
single-group case, encoding costs $O(n)$ over the entire input
and there is no interface work. The micro TNFA computations cost
$O(m/x)$ per block in both cases. The working space is $O(m)$,
including state masks and the interface matrices, which use
$O(y)$ words per group. Initialize the state set as in
Section~\ref{subsec:fast-full}. Process consecutive blocks of length
$y$ and a possible shorter final block. Test whether the final
state set contains the accepting state of $N$. The total matching time is
\[
 O\left(\ceil{n/y}\frac{m}{x}+m+n\right)
   = O\left(\frac{nm\,\lg M}{M^2}+m+n\right).
\]
If $M=L$, this is the claimed matching term. If $M=m<L$, then
$nm\,\lg M/M^2=O(n)$. Adding preprocessing and using $t\geq m$
gives the theorem.
\end{proof}

For fixed $\eps>0$ and $t\geq m^{1+\eps}$, we have
$\lg(t/m)=\Theta(\lg t)$ and
$\lg\lg(t/m)=\Theta(\lg\lg t)$ for sufficiently large $t$.
In this range, the time bound simplifies to
$O(nm\lg\lg t/\lg^2 t+t+n)$.

\bibliographystyle{alpha}
\bibliography{master}

\end{document}